\pdfoutput=1

\documentclass[10pt,letter,twocolumn]{article}
\usepackage{f1000_styles}

\usepackage[super,comma,sort]{natbib}
\usepackage[colorlinks=true,urlcolor=blue]{hyperref}

\usepackage{amsmath}

\newcommand*{\T}{T}
\newcommand*{\A}{\mathcal{A}}

\newcommand*{\Tz}{{\T}_z}

\newcommand*{\qz}{q_z}

\newcommand*{\qh}{\hat{q}}
\newcommand*{\qt}{\tilde{q}}

\newcommand*{\dz}{\dd z}
\newcommand*{\dq}{\dd q}
\newcommand*{\dw}{\dd w}

\newcommand*{\dT}{\dd \T}
\newcommand*{\Gpz}{\Gp_z}
\newcommand*{\dGpz}{\dd\Gpz}
\newcommand*{\dGp}{\dd\Gp}
\newcommand*{\qdGp}{q\,\dd\Gp}

\newcommand*{\xdot}{\dot{x}}
\newcommand*{\Gz}{\xi}
\newcommand*{\tldh}{\tilde{h}}

\newcommand*{\ave}[2]{\angb{#1}_{#2}}

\newcommand*{\Ga}{\alpha}
\newcommand*{\Gb}{\beta}

\newcommand*{\Gg}{\gamma}
\newcommand*{\GG}{\Gamma}

\newcommand*{\Gl}{\lambda}

\newcommand*{\Gp}{\psi}

\newcommand*{\angb}[1]{\left\langle#1\right\rangle}
\newcommand*{\lrp}[1]{\left(#1\right)}

\newcommand*{\dd}{\textrm{d}}

\newcommand*{\EEq}[1]{Eqn~\ref{eq:#1}}
\newcommand*{\Eq}[1]{eqn~\ref{eq:#1}}

\newcount\BoxNum \BoxNum 1\relax
\makeatletter
\newcommand*{\boxlabel}[1]{%
  \protected@write \@auxout {}{\string \newlabel {box:#1}{{\the\BoxNum}}{}}%
  \advance\BoxNum 1\relax}
\makeatother

\newcommand*{\boldrule}{\hrule height 1.2pt}
\newcommand*{\noterule}{\medskip\boldrule\medskip}	

\def\citeyear{\citep}
\def\autocite{\citep}
\def\textcite{\citet}

\begin{document}

\title{Invariant death}
\author[1]{Steven A.~Frank}
\affil[1]{Department of Ecology and Evolutionary Biology, University of California, Irvine, CA 92697--2525 USA, safrank@uci.edu}

\maketitle
\thispagestyle{fancy}

\parskip=4pt

\begin{abstract}

In nematodes, environmental or physiological perturbations alter death's scaling of time. In human cancer, genetic perturbations alter death's curvature of time. Those changes in scale and curvature follow the constraining contours of death's invariant geometry. I show that the constraints arise from a fundamental extension to the theories of randomness, invariance and scale. A generalized Gompertz law follows. The constraints imposed by the invariant Gompertz geometry explain the tendency of perturbations to stretch or bend death's scaling of time. Variability in death rate arises from a combination of constraining universal laws and particular biological processes.

\end{abstract}


\vskip0.5in
\noterule
Preprint of published version: Frank, S. A. 2016. Invariant \hfil\break death. F1000Research 5:2076,  \hfil\break \href{http://dx.doi.org/10.12688/f1000research.9456.1}{doi:10.12688/f1000research.9456.1}. \break Published under a Creative Commons \href{https://creativecommons.org/licenses/by/4.0/}{CC BY 4.0} license.
\smallskip
\noterule

\clearpage

\let\origquote\quote
\let\endorigquote\endquote
\renewenvironment{quote}{%
  \sffamily
  \origquote
}%
{\endorigquote}


\section*{Introduction}

The coil of a snail's shell expresses the duality of constraint and process. The logarithmic spiral of growth constrains overall form. Particular snails modulate the process of shell deposition, varying the parameters of the logarithmic spiral. To interpret the variety of snail shells, one must recognize the interplay between broad geometric constraint and the special modulating processes of individual types\autocite{thompson92on-growth}.

The pattern of death in populations follows the same duality of invariant geometric constraint and modulating process. The invariant geometry of death's curve arises from the intrinsic order of large samples\autocite{feynman98statistical,gnedenko68limit}. A large sample erases underlying randomness, preserving only invariant aggregate values\autocite{jaynes03probability}. 

I extend the large-sample concept to clarify the invariant geometry of death. I then illustrate the role of particular biological processes in modulating death's curve: the stretch of death's time in nematode response to physiological perturbation\autocite{stroustrup16the-temporal} and the curvature of cancer's time in response to genetic perturbation\autocite{frank05age-specific,frank07dynamics}. The consequences of particular biological perturbations can only be understood within the geometry that constrains change to follow invariant contours.

To restate the puzzle: How can we relate small-scale molecular and physiological process to population consequence? The problem remains unsolved. \textcite{finch16constant} emphasized: ``A key question is how to connect $\ldots$ [linear] aging processes to the exponential rates of accelerating mortality that set life spans. $\ldots$ Although we can readily assess molecular aging, such biomarkers of aging are rarely robust as predictors of individual morbidity and mortality risk in populations.''

\section*{Randomness and invariance}

I begin with the relation between small-scale randomness and large-scale order. The classical theory derives from the principles of statistical mechanics\autocite{feynman98statistical}, later developed through aspects of entropy and information\autocite{jaynes03probability,presse13principles}. Here, I briefly summarize my own extension of classical results based on geometric principles of invariant measurement and scale\autocite{frank10measurement,frank11a-simple,frank14how-to-read,frank16common}. I then show how the abstract geometry constrains the relation between biological process and the pattern of death's curve. 

To understand the probability of dying at a particular age, we begin with the geometry of probability patterns\autocite{frank16common}. For an underlying quantity, $z$, the probability of observing a value near to $z$ is the rectangular area with height $\qz$, width $\dGpz$, and area $\qdGp$. A probability pattern is a curve with coordinates $\lrp{\Gpz,\qz}$ defined parametrically with respect to $z$. For the curve of death, the input, $z$, may be age or time. 

Two invariances constrain the geometry of probability curves. First, total probability is invariantly one. Invariant total probability implies that the height of the probability curve has a natural exponential expression\autocite{frank16common}
\begin{equation}\label{eq:shift}
  \qz = k_a e^{-\Gl(a + \Tz)} = k e^{-\Gl\Tz},
\end{equation}
in which $k=k_a e^{\Gl a}$ remains constant for any $a$ to satisfy the requirement that total probability is invariant. The exponential form for the height of the probability curve, $\qz$, implies that the probability curve remains invariant to a shift of the fundamental metric, $\Tz$ (see \textcite{frank16common}). 

In general, we seek metrics for which it does not matter where we set our zero reference point. In geometry, a circle shifted in space retains its invariant form. Similarly, proper geometric scaling for probability patterns is shift invariant. In terms of death, any transformation of time, $z$, into a fundamental time metric for probability pattern, $\Tz$, must measure time such that a shift $a+\Tz$ does not alter death's curve. That shift-invariant requirement leads to the exponential expression\autocite{frank16common} in \Eq{shift}.

The second key invariance is that a uniform stretching or shrinking of the fundamental metric does not alter probability pattern\autocite{frank16common}
\begin{equation}\label{eq:stretch}
  \qz = k e^{-\Gl_bb\Tz} = k e^{-\Gl\Tz},
\end{equation}
in which $\Gl=\Gl_b b$ remains constant for any $b$, causing the probability pattern to be invariant to stretch of $\Tz$. Stretch invariance is equivalent to invariance of $\Gl\ave{\T}{\Gp}$, the value of $\Gl$ multiplied by the average value of $\T$ when probability, $\qdGp$, is measured on the scale, $\Gp$ (see \textcite{frank16common}).

To summarize, probability curves remain invariant to shift and stretch of the fundamental metric, $\Tz$, such that
\begin{equation}\label{eq:affine}
  \T\mapsto a+b\T \sim T,
\end{equation}
in which `$\sim$' means invariant with respect to shift and stretch. In geometry, invariance with respect to shift and stretch is affine invariance.

Affine invariance leads to probability pattern described by a sequence of rectangular areas
\begin{equation*}
  q\,\dGp = ke^{-\Gl\T}\dGp,
\end{equation*}
in which $k$ and $\Gl$ are constants that adjust to satisfy, respectively, invariant total probability and invariant average value, $\Gl\ave{\T}{\Gp}$. Many different approaches and interpretations all arrive at this same basic form. 

\section*{Consequences of affine invariance}

Here, by emphasizing the fundamental invariances, we can take the next key step in understanding the geometry of probability patterns and the curves of death. In particular, each successive application of the affine transformation (\Eq{affine}) to $\T$ leaves the probability pattern unchanged, defining an invariant group of metrics\autocite{frank11a-simple}
\begin{equation}\label{eq:Tuniversal}
  \T= \frac{1}{\Gb}\lrp{e^{\Gb w}-1} \sim e^{\Gb w},
\end{equation}
with $\Gb\rightarrow 0$ implying $\T\rightarrow w$. Here, $w(z)$ is a scale for the underlying values, $z$, such that a shift in that scale, $w\mapsto\Ga+w$, only changes $\T$ by a constant multiple, and therefore does not change the probability pattern. 

To find the proper metric, $\T$, for a particular probability pattern, we only need to find the proper base scale $w$ for which the probability pattern is shift invariant. If, for example, $z$ is time or age, then we only need to discover the scaling, $w(z)$, for which 
\begin{equation}\label{eq:common}
  q\,\dGp = ke^{-\Gl e^{\Gb w}}\dGp
\end{equation}
is invariant to a shift in $w$, when allowing adjustment of $\Gl$. When $\Gb\rightarrow0$, then $q\rightarrow e^{-\Gl w}$.

\EEq{common} expresses the abstract form of common probability patterns\autocite{frank11a-simple}. The abstraction does not specify the two key scaling relations $\Gp(z)$ and $w(z)$ that define the coordinates of the parametric probability curve $\lrp{\Gp,q}$ with respect to $z$. However, the invariances that define the geometry do impose strong constraints, leading to a limited set of forms for almost all of the commonly observed probability patterns\autocite{frank11a-simple,frank14how-to-read,frank16common}.

We have two scaling relations $\Gp$ and $w$, but only a single parametric probability curve $\lrp{\Gp,q}$ with associated probability $\qdGp$ in each increment. Thus, many different scales can express the same probability pattern. For each application, there is often a natural scale that has a simple, understandable form for its scaling relations. 

\section*{The invariant ticking of death's clock}

A natural scale corresponds to an additional invariance with a simple interpretation. That additional invariance sets the underlying metric for the pair of scaling relations. For death, we can set the probability of dying to be invariant in each increment of the scale, $\dGp$, so that $\Gp$ represents the uniform metric of mortality---the invariant ticking of death's clock. This uniform metric extends the theory of extreme values and time to failure\autocite{embrechts97modeling,kotz00extreme,coles01an-introduction,gumbel04statistics} to a more abstract and general understanding of the invariances that shape all of the common probability patterns\autocite{frank11a-simple,frank14how-to-read,frank16common}.

Invariant probability in each increment can be written as $\qdGp=-\dq$ and thus $\Gp=-\log q$, in which $\dq$ is a constant incremental fraction of the total probability. I use a minus sign as a convention to express the total probability as declining with an increase in $\Gp$. 

With regard to dying, we may think of the total probability of being alive as declining by a constant increment of death, $-\dq$, in each increment $\dGp$.  In classic epidemiology, this definition of $q$ would be expressed as $q(z)\equiv S(z)$, in which $S(z)$ is the probability of survival to time $z$. However, it is important to consider the classic definition as a special case of the deeper abstract geometry, which leads to a more general understanding of the constraints that shape death's curve.

\section*{Universal Gompertz geometry}

Given the exponential form for $\qz$ in \Eq{stretch}, a constant probability $\qdGp$ in each increment requires $\dGp=\dT$. Using the general form of $\T$ in \Eq{Tuniversal}, we have $\dT= e^{\Gb w}\dw=\T'\dw$, in which $\T'>0$ is the derivative of $\T$ with respect to $w$. With $\qh\,\dw=q\,\dT$ for the constant probability in each increment, we have
\begin{equation}\label{eq:gompertz}
  \qh\,\dw = k e^{\log\T' - \Gl\T}\dw= k e^{\Gb w - \Gl e^{\Gb w}}\dw.
\end{equation}
This probability pattern is expressed on the scale $w$, in which $w$ defines the natural shift-invariant metric. In other words, for some underlying observable value $z$, such as time or age, $w(z)$ transforms $z$ to a scale, $w$, that expresses an invariant total probability $\qh\,\dw$ in each increment, and for which shifts in the scale $w\mapsto\Ga+w$, do not change the probability pattern. 

The probability pattern in \Eq{gompertz} has the familiar Gompertz form. I derived that form solely from a few simple geometric invariances. The simple invariances elevate the generalized Gompertzian form to a universal geometric principle for probability patterns\autocite{frank11a-simple,frank16common}. By contrast, the Gompertzian pattern is usually derived from descriptive statistics or from particular assumptions about processes of failure or growth. 

\section*{Pattern on the observed scale}

We may express the probability pattern on the scale of the underlying observable value, $z$. For that scale, $\dw=w'\dz$, in which $w'>0$ is the derivative of $w$ with respect $z$. The abstract Gompertzian geometry in $w$ becomes the explicit form with respect to the directly measured value $z$ as
\begin{equation}\label{eq:gompertzz}
  \qt\,\dz = k e^{\log w'\T' - \Gl\T}\dz= k w' e^{\Gb w - \Gl e^{\Gb w}}\dz.
\end{equation}

\section*{The hazard of death}

Only living individuals can die. Thus, the hazard of death is the probability of dying in an incremental metric of time divided by the probability of being alive. The incremental metric scale, $\Gpz$, transforms the observed value, $z$, which may be time or age, into the abstract incremental scale, $\dGp$. The abstract expression for the hazard of dying in an increment $\dGp$ is
\begin{equation}\label{eq:hGp}
  h(\Gp)\dGp = \frac{\qdGp}{1-\int \qdGp}.
\end{equation}
In each increment, the probability of dying is $\qdGp$. The integral in the bottom is the sum of the probabilities of dying over the period from a starting point until the current period, in which the time metric is described by $\Gp(z)$. 

Three different metrics transform the observable time input, or other measurable input, $z$, into the scale of analysis: $\T$, $w$, and $z$ itself. Those three metrics yield three equivalent forms for the hazard, each emphasizing different aspects of the underlying geometric invariances 
\begin{align}
  h(\T)\dT &= \Gl\dT \propto\dT \label{eq:hT} \\[3pt]
  \hat{h}(w)\dw &= \Gl\T'\dw \propto e^{\Gb w}\dw \label{eq:hw}\\[3pt]
  \tldh(z)\dz &= \Gl w'\T'\dz \propto w'e^{\Gb w}\dz  \label{eq:hz}
\end{align}
in which `$\propto$' denotes proportionality. The top form expresses the most general and abstract invariance of death. By transforming time into a general metric, $z\mapsto \T$, the hazard is invariantly $\Gl$ in each increment of the metric, $\dT$. The metric $\T$ defines the scale on which the probability pattern is invariant to the affine transformation $\T\mapsto a + b\T$. 

We know the scale of death's curve when we can transform our underlying observation, $z$, such as age, to the affine-invariant scale, $\T$. Often, adding a constant to age or multiplying age by a constant, $z\mapsto a+bz$, changes the pattern of death's curve, so using age itself as the metric is usually not sufficient. We must find some transformation of age.

The middle expression in \Eq{hw} describes the generalized Gompertzian geometry in the most direct way. In this case, when we transform $z\mapsto w$, changing an observation such as age, $z$, to the metric, $w(z)$, we only require that death's curve be invariant to a shift, $w\mapsto a+w$. 

\section*{The force of death and the curvature of time}

We are partitioning the scaling of death's curve into two steps, $z\mapsto w\mapsto \T$. Once we have the shift-invariant scaling of time, $w$, then $\T=e^{\Gb w}$ changes $w$ into the ultimate affine-invariant scaling, $\T$. To make that last change, we need to know $\Gb$, which is 
\begin{equation}\label{eq:beta}
  \Gb=\frac{\T''}{\T'} = \frac{\dd\log\hat{h}}{\dw},
\end{equation}
in which each prime denotes the derivative with respect to $w$. This value of $\Gb$ defines the curvature for the geometry of death and time.  Once we have the shift-invariant scaling for time, $z\mapsto w$, we can consider death's invariant curvature in the transformation $w\mapsto\T$. 

The expression $\T''$ is the acceleration, or absolute curvature, of $\T$. The expression $\T'$ is the rate or velocity at which $\T$ is changing. Thus, $\T''/\T'$ can be thought of as the acceleration relative to the velocity. 

Acceleration, curvature and force are ultimately equivalent. In terms of death, for a given velocity or rate at a particular age, $\T'$, the value of $\Gb$ is the relative force that bends death's curve. The bending of death's curve may also be described as \begin{equation}\label{eq:accel}
  \A = \frac{\dd\log\hat{h}}{\dw} = \frac{\hat{h}'}{\hat{h}},
\end{equation}
which is the change in the hazard of death relative to the current hazard. The hazard, $\hat{h}$, is the rate, or velocity, of death on the scale $w$. Thus, the change in relative velocity, $\A$, describes the acceleration of mortality in terms of the relative bending of death's rate. 

The invariant geometry of death's curve in \Eq{beta} may be expressed as a balance, $\Gb - \A=0$, between force and acceleration. That balance is roughly analogous to Newton's second law of motion, $F=m\A$, relating force to acceleration.

\section*{Inference}

The invariant geometry does not tell us the form for the shift-invariant scaling of death's time, $w$, or the value of the invariant force, $\Gb$, that bends death's curve. However, the invariances strongly constrain the likely form of death's curve and the meaningful metrics of death's time. Importantly, these expressions allow us to transform data about rates or motions into expressions that emphasize force and causal interpretations\autocite{lanczos86the-variational,frank15dalemberts}. In biology, we rarely can predict trajectories. Instead, we focus on interpreting the changes in observed trajectories with respect to hypothesized forces\autocite{frank07dynamics,frank14the-inductive}. 

The abstract geometry is correct unto itself. In application, the geometry provides a tool that we may use for particular problems. A tool is neither right nor wrong. Instead, a tool is helpful or not according to its aid in providing insight. Below I discuss some examples. A few comments prepare for the discussion.

If we knew the correct scaling for age, $w(z)$, then within that frame of reference, the force, $\Gb$, and acceleration, $\A$, of mortality would be constant with respect to $w$. Thus, the frame of reference, $w$, provides valuable insight. However, $w$ may turn out to be a weirdly nonlinear scaling of measured time, $z$, in which the form of $w$ is difficult to determine directly. In practice, we can derive $w$ from \Eq{hz} by relating the hazard, $\tldh(z)$, to $w$ by
\begin{equation}\label{eq:wscale}
  \Gb w = \log \int\tldh\dz,
\end{equation}
or $\Tz\sim\int\tldh\dz$, the affine similarity of $\Tz$ to the accumulated hazard on the $z$ scale.

I now discuss the time scaling of mortality in nematodes and cancer. I consider these applications only to illustrate general aspects of mortality's temporal geometry. See \textcite{stroustrup16the-temporal} for details about nematodes and \textcite{frank07dynamics} for details about cancer.

\section*{Nematode mortality and the stretch of time}

\textcite{stroustrup16the-temporal} conclude from their study of nematode mortality:
\begin{quote}
[W]e observe that interventions as diverse as changes in diet, temperature, exposure to oxidative stress, and disruption of [various] genes $\ldots$ all alter lifespan distributions by an apparent stretching or shrinking of time. To produce such temporal scaling, each intervention must alter to the same extent throughout adult life all physiological determinants of the risk of death. 
\end{quote}
I begin with the apparent \textit{stretching or shrinking of time}. I will arrive at the same description of the nematode mortality pattern as given by \textcite{stroustrup16the-temporal}, but framed within my more general understanding of mortality's invariant geometry. From that broader perspective, the observed stretching or shrinking of time in the nematode study can be seen as a special case of the various temporal deformations that arise with respect to mortality's invariant scale.

The perspective of my general framing calls into question the second conclusion that \textit{each intervention must alter to the same extent throughout adult life all physiological determinants of the risk of death}. I present a simple counterexample consistent with the observed patterns. My counterexample may not be the correct description of process in nematode mortality. The counterexample does, however, emphasize key aspects of the logic by which we must evaluate the relations between pattern and process in mortality. 

My framework analyzes the sequence of transformations $z\mapsto w \mapsto\T$. The initial input, $z$, typically represents what we measure, such as a standard description of time or age. We then seek a transformation, $w(z)$, such that the parametric curve, $\lrp{w,\hat{q}}$, for observed or assumed probability pattern is shift-invariant with respect to $w$ (\Eq{gompertz}). In other words, the shift $w\mapsto \Ga + w$ does not alter the probability curve. When we find the shift-invariant scale for $w$, we have an expression for the probability pattern in terms of the Gompertzian geometry of \Eq{gompertz}. 

A probability pattern that remains the same except for a constant stretching or shrinking of time corresponds to $w(z) = \log z$, because a constant stretch or shrink of time by $a=e^\Ga$ yields $w(az) = \Ga+w(z)$. If we express the associated parametric probability curve as the relation between time and probability, $\lrp{z,\tilde{q}}$, as in \Eq{gompertzz} with $w=\log z$, we obtain a curve that is invariant to a constant stretching or shrinking of the temporal scale, $z$, as
\begin{equation*}
  \tilde{q} \propto z^{\Gb-1}e^{-\Gl z^\Gb}, 
\end{equation*}
in which the parameter $\Gl$ and the constant of proportionality both adjust to cancel any stretch or shrink of time by $a>0$ (see \textcite{frank16common}). This curve is the Fr{\'e}chet probability distribution, corresponding to the power law hazard in \Eq{hz} as 
\begin{equation*}
  \tldh \propto z^{\Gb-1}.
\end{equation*}
\textcite{stroustrup16the-temporal} concluded that the Fr{\'e}chet distribution is the best overall match to their nematode studies. However, they invoked the Gompertz-Fr{\'e}chet family of distributions by appeal to traditional epidemiology and by appeal to the general form of extreme value distributions for failure times. By contrast, I derived those distributions simply as the inevitable consequence of basic assumptions about the invariant geometry of meaningful scales\autocite{frank16common}. 

\section*{The deformation of death's time}

\textcite{stroustrup16the-temporal} discussed the stretching or shrinking of death's time by a single constant value. My framework generalizes the deformation of time in relation to death. We begin with $\T$, the universal frame of reference for the scaling of death's time. On the temporal scale, $\T$, the hazard of death, $h(\T)$, remains constant at all times (\Eq{hT}). Thus, $\T$ represents the invariant ticking of mortality's clock. 

Given that universal frame of reference for time, we may then consider other temporal scales in terms of the way in which they deform the invariant frame of reference. In this case, we work inversely, by starting with $\T$ in \Eq{Tuniversal}, and then inferring the deformations with respect to the underlying scale of description, $z$. We can then think of the shape of the curve $\lrp{\Tz,z}$ as describing how measured time, $z$, is deformed in relation to the universal invariant scale of mortality's time, $\Tz$. 

Ideally, we first infer the shift-invariant scale, $w(z)$, and then use $w$ in \Eq{Tuniversal} to determine the relation between $\T$ and $z$. In the nematode case, $w(z)=\log z$ achieved shift invariance. Thus $\Tz\sim e^{\Gb w} = z^\Gb$. The power law curve $\lrp{z^\Gb,z}$, with curvature determined by $\Gb$, describes the deformation of time. The different experimental treatments did not significantly alter the curvature associated with $\Gb$.

We can relate increments of the measured input, $\dz$, to increments of mortality's universal measure, $\dT$, by starting with \Eq{hz} as $\tldh\propto\dT/\dz$, and then writing 
\begin{equation}\label{eq:zscale}
  \dz\propto\dT/\tldh.
\end{equation}

For a case such as the nematodes in which $\Tz\sim z^\Gb$, the measured temporal increments, $\dz$, scale in relation to the universal temporal frame as $\dz\propto\dT/z^{\Gb-1}$. For $\Gb>1$, measured temporal increments, $\dz$, shrink as time passes relative to the constant ticking of mortality's clock at $\dT$. When we think of $\dT$ as mortality's constant temporal frame of reference, then the deformation of measured time is
\begin{equation*}
  \dz\propto\frac{1}{z^{\Gb-1}}.
\end{equation*}
The shrinking of measured time corresponds to the increase in the rate of measured mortality, in other words, the same amount of mortality, $\dT$, is squeezed into smaller temporal increments, $\dz$, increasing the density of mortality per measured unit.

In other cases, the relation of measured inputs, $z$, to mortality's universal scale, $\Tz$, will have different functional forms. Those different functional forms may correspond to non-uniform stretching and shrinking of the observed temporal scale at different magnitudes of $z$ relative to the universal frame of reference for mortality on the scale $\Tz$. If possible, we first infer the shift-invariant scale, $w(z)$, for example by \Eq{wscale}, and then use $w$ to determine the relation between $\T$ and $z$, as in the nematode example. However, in practice, it may be easier to go directly from the invariant clock, $\T$, to the deformed time scale, $z$, by using the relation $\dz\propto\dT/\tldh$. The following critique of the conclusions by \textcite{stroustrup16the-temporal} about nematode mortality provides an example. 

\section*{Invariant pattern and underlying process}

\textcite{stroustrup16the-temporal} claimed that all physiological determinants of the risk of death change in the same way with each experimental intervention. I present simple counterexamples. Although my counterexamples may not describe the true underlying process, they do highlight two important points. First, commonly observed patterns often express invariances that are consistent with many alternative underlying processes\autocite{frank09the-common,frank14generative}. Second, consideration of the alternative processes with the same observable invariances leads to testable predictions about the underlying causal processes.

In these examples, suppose that death follows a multistage process, as is often discussed in cancer progression\autocite{armitage54the-age-distribution}. Following \textcite[p.~98]{frank07dynamics}, we may write the dynamics of progression toward mortality as a sequence of transitions
\begin{align*}
		\xdot_0(z) &= -u_0x_0(z) \\
		\xdot_i(z) &= u_{i-1}x_{i-1}(z) - u_ix_i(z) 
				\mskip40mu  i=2,\dots,n-1  \\
		\xdot_n(z) &= u_{n-1}x_{n-1}(z),
\end{align*}
where $x_i(z)$ is the fraction of the initial population born at time $z=0$ that is in stage $i$ at measured time, $z$. Assume that when the cohort is born, at $z=0$, all individuals are in stage 0, that is, $x_0(0)=1$, and the fraction of individuals in other stages is zero.  

As time passes, some individuals move into later stages of progression toward death.  The rate of transition from stage $i$ to stage $i+1$ is $u_i$.  The $\xdot$'s are the derivatives of $x$ with respect to $z$. Death occurs when individuals transition into stage $n$. A fraction $x_n(z)$ of individuals has died at time $z$, and the rate of death at time $z$ is $\xdot_n(z)\equiv\tilde{q}$, in which $\tilde{q}$ has the probability interpretation of \Eq{gompertzz}. 

If the transition rates are constant and equal, $u_i=u$ for all $i$, then we can obtain an explicit solution for the multistage model \autocite{frank04a-multistage}.  This solution provides a special case that helps to interpret more complex assumptions.  The solution is $x_i(z) = e^{-uz}(uz)^i/i!$ for $i=0,\dots,n-1$, with the initial condition that $x_0(0)=1$ and $x_i(0)=0$ for $i>0$.  Note that the $x_i(z)$ follow the Poisson distribution for the probability of observing $i$ events when the expected number of events is $uz$.

In the multistage model above, the derivative of $x_n(z)$ is given by $\xdot_n(z)=ux_{n-1}(z)$.  From the solution for $x_{n-1}(z)$, we have 
\begin{equation*}
  \tilde{q}=\xdot_n(z)=ue^{-uz}(uz)^{n-1}/n-1!,
\end{equation*}
which is a gamma probability distribution. One can think of the gamma distribution as the waiting time for the $n$th event, in which each event occurs at constant rate, $u$. However, many other processes lead to the same gamma distribution.
 
Age-specific incidence is the hazard\autocite{frank07dynamics}
\begin{equation}\label{eq:hgamma}
	\tldh(z) = {{\xdot_n(z)}\over{1-x_n(z)}} 
			= \frac{u(uz)^{n-1}/n-1!}{\sum_{i=0}^{n-1}(uz)^i/i!}.
\end{equation}
We can express the scaling of measured time, $\dz$, relative to the constant ticking of mortality's time, $\dT$, from \Eq{zscale}, by taking $\dT$ as constant and thus
\begin{equation} \label{eq:zgamma}
\begin{aligned}
  \dz\propto\frac{1}{(uz)^{n-1}}\Bigg(1+uz+\frac{(uz)^2}{2}&+\dots\\
  	&+\frac{(uz)^{n-1}}{n-1!}\Bigg).
\end{aligned}
\end{equation}

\section*{Simultaneity and temporal deformation}

When measured time, $z$, is small, during the initial period of the process, the deformation of time is approximately the same as the Fr{\'e}chet pattern, $\dz\propto 1/z^{n-1}$. This deformation in the gamma process describes the force of simultaneity. Early in the process, all components that protect against mortality remain in the initial working state. Thus, mortality requires the nearly simultaneous failure of $n$ independent events, which creates a force that deforms the constant ticking of mortality's clock by rescaling the measured increments, $\dz$. 

As measured time increases, the increments $\dz$ shrink, compressing the same amount of mortality, $\dT$, into smaller measured temporal increments. As $z$ becomes larger, the increments $\dz$ in \Eq{zgamma} shrink less, because of the reduced force of simultaneity that deforms mortality's constant clock. With larger $z$, the higher-power terms of the sum increasingly dominate, until the largest power term dominates and $\dz$ then ticks at a constant rate, with $\dz\propto\dT$. 

The changing deformation of $\dz$ and the associated force of mortality can be thought of roughly as follows\autocite{frank07dynamics}. Early in the gamma process, mortality requires the nearly simultaneous failure of $n$ independent events, creating a force of simultaneity such that $\dz\propto1/z^{n-1}$. As time passes, many individuals suffer failure of some of the $n$ processes, leaving in aggregate the equivalent of $n-1$ remaining protective components, and a force of simultaneity such that, approximately, $\dz\propto1/z^{n-2}$. As more time passes, additional components fail, and the remaining force of simultaneity diminishes, until eventually only one protective component remains for those still alive, at which point $\dz$ then ticks at a constant rate, so that $\dz\propto\dT$.

We may also express the scaling of time on the shift-invariant Gompertzian scale, $w$, in which $\Gb$ is a relative measure of the acceleration of mortality (\Eq{beta}), by using the general expression in \Eq{wscale} and the specific form of $\tldh$ in \Eq{hgamma} to yield
\begin{equation*}
  \Gb w = \log \int\tldh\dz = \log \log \GG(n,uz)^{-1},
\end{equation*}
in which $\GG$ is the incomplete gamma function.

\section*{Alternative models of nematode mortality}

With this understanding of the gamma process, we can consider alternative interpretations of the nematode data\autocite{stroustrup16the-temporal}. I present these alternatives to illustrate the logic of mortality's temporal scaling and the potential relation to underlying process. The data do not provide information about whether or not these alternative interpretations are the correct description of nematode mortality. The point here is that these alternatives, or some other structurally similar alternative, might be correct, and therefore the strong conclusions of the original article may be false.

To repeat the key conclusion from \textcite{stroustrup16the-temporal}, \textit{each intervention must alter to the same extent throughout adult life all physiological determinants of the risk of death}.

That conclusion is true for the simple gamma process, as summarized by \Eq{zgamma}. In that equation, the value of $u$ represents the rate at which each of the $n$ processes fails and contributes to overall mortality. If we substitute $uz\mapsto\Gz$, then the scaling of measured time, expressed as $\dz\propto\dd\Gz$, changes only by a constant of proportionality as the rate, $u$, changes.

I now consider two variations on the underlying gamma process for mortality. Each of these variations leads to a constant rescaling of time, $\dz$. However, that constant rescaling arises from underlying processes of mortality that change in different ways in response to perturbations. These examples show that the constant rescaling of time does not imply that an intervention alters to the same extent throughout adult life all physiological determinants of the risk of death.

The first example considers two distinct sets of underlying processes that influence mortality, each set composed of $n$ processes. Mortality occurs only after the failure of all $2n$ processes. Before experimental perturbation, one set of $n$ processes has a relatively slow failure rate per process of $u$. The other set of $n$ processes has a relatively fast failure rate of $u'\gg u$. 

In this case, the fast processes will all tend to fail early in life, almost always before all of the slow processes fail. Thus, the fast processes have little influence on mortality. The mortality rate will closely follow the gamma process with $n$ steps, each step at rate $u$, as analyzed above\autocite{frank07dynamics}.

Now suppose that an intervention influences all of the slow steps but none of the fast steps. The intervention changes the previously slow rate processes into fast processes, $u\mapsto u''\gg u'$. After intervention, the mortality rate will closely follow the gamma process with $n$ steps, with each step at rate $u'$. The mortality pattern remains unchanged, except for a constant rescaling of time. However, the underlying physiological processes that determine mortality have changed completely. Previously unimportant rate processes with respect to mortality now completely dominate, and previously important rate processes no longer influence mortality.

The second example considers a set of $n$ underlying processes that influence mortality. Each process has a different failure rate of $u_i$ for $i=1,\dots,n$, with $u_i < u_{i+1}$. As before, mortality occurs only after the failure of all $n$ processes. \textcite{frank07dynamics} presented numerical studies for this heterogeneous rate process model. Typically, the faster rate processes fail early in life and have relatively little influence. The slower processes dominate the overall temporal pattern. 

With $n$ equal rate processes, the curvature declines with time, as in \Eq{zgamma}. With a heterogeneous set of rate processes, the curvature tends to decline more quickly as the fastest processes fail earlier, typically leaving a progressively smaller set of remaining protective mechanisms as time passes, reducing the force of simultaneity.

Now suppose that the heterogeneous set of $n$ processes has a simple hierarchy of rates, such that $u_{i+1} = \Gg u_i$, in which $\Gg>1$ is the factor by which each rate increases relative to its slower neighbor. If the only effect of an experimental intervention is abrogation of the slowest process, $u_1$, then the hierarchy is effectively altered only by multiplying each rate in the set by $\Gg$, because the fastest processes typically have almost no influence on pattern. 

Once again, the overall mortality pattern will change only by a constant rescaling of time, even though the underlying physiological processes have changed significantly with respect to their influence on mortality. In this case, the most important process that limited mortality before intervention was in effect knocked out after intervention, whereas all other processes did not change.

\section*{Different processes lead to same invariances}

The actual biology of nematodes will, of course, not follow exactly either of these two example cases. The examples do show, however, that a constant rescaling of measured time for mortality can arise by heterogeneous changes in the underlying physiological determinants of the risk of death.

How should we interpret the match between variants of the multistage gamma process model and the observed scaling of nematode mortality? The correct view is that the invariances expressed by the gamma model are approximately the same as the invariances that arise by the true physiological processes. Those invariances dominate the shape of the observed patterns. The examples of the gamma models are helpful, because they show the sorts of underlying processes that generate the required invariances. 

Ultimately, the theoretical challenge is to understand the full set of underlying processes that lead to the same invariances and thus the same observed pattern\autocite{frank14generative}. The empirical challenge is, of course, to figure out which particular processes occur in each particular case. Success in the empirical challenge will likely depend on further progress on the theoretical challenge, because the theoretical frame strongly influences how one goes about solving the empirical problem.

\section*{Cancer incidence and the curvature of time}

I now turn to genetic knockouts in cancer that change the curvature of time. Cancer incidence often follows a pattern roughly consistent with a multistage gamma process\autocite{armitage54the-age-distribution,frank07dynamics}. Again, that match does not mean that the underlying physiological processes truly follow the assumptions of the gamma model. Instead, the correct view is that the invariances expressed by the gamma model are approximately the same as the invariances that arise by the true physiological processes. 

Consider the simplest gamma process, in which cancer arises only after $n$ protective mechanisms fail. Each mechanism fails at the same rate, $u$. I gave the explicit solution for that process earlier. In interpreting that solution for cancer, it is important to note an essential distinction between mortality and cancer incidence. 

Everyone dies but only a small fraction of individuals develop a particular form of cancer. Thus, we must analyze mortality by running the measured time, or age, $z$, out to a large enough value so that the cumulative probability of dying approaches one. In the gamma models above, that means letting $uz$ increase significantly above one. By contrast, if only a small fraction of the population develops cancer before dying of other causes, then we must run $z$ only up to a time at which the cumulative probability of cancer remains small. That limit on total incidence typically means capping $uz$ below one. 

With a small maximum value of $uz$, the age-specific hazard simplifies approximately to $\tldh\propto z^{n-1}$, and the scaling of measured time simplifies to $\dz\propto1/z^{n-1}$. Those scalings match the Fr{\'e}chet model when we equate the curvature of time, $\Gb$, with the number of steps, $n$, and we interpret force and curvature with respect to the shift-invariant scaling of time, $w(z)=\log z$. 

We can think of $n=\Gb$ as the force imposed on the logarithmic time scale, $\log z$, caused by the requirement for the nearly simultaneous failure of $n$ protective processes. The greater $n$, the greater the protective force, and the greater the bending of observed time relative to cancer's invariant clock, ticking in increments of $\dT$. 

All of that may seem to be a very abstract theory in relation to the actual physiological processes of cancer. However, certain empirical studies suggest that the simple geometric theory of cancer's time does in fact capture key aspects of cancer's real physiology and genetics.  In particular, certain inherited genetic mutations correspond almost exactly to the predicted theoretical change in the force of simultaneity and the temporal curvature of incidence. 

If a mutational knockout reduces the number of protective mechanisms by one, such that $n\mapsto n-1$, then the approximate pattern of incidence changes from $\tldh\propto z^{n-1}$ to $\tldh\propto z^{n-2}$. In other words, the force and associated curvature, $\Gb$, is reduced by one. 

Two classic studies of cancer incidence made exactly that comparison. \textcite{ashley69colonic} compared colorectal cancer incidence between groups with and without an inherited mutation that predisposes to the disease. Similarly, \textcite{knudson71mutation} compared retinoblastoma incidence between groups with and without a predisposing inherited mutation. 

I analyzed those same cancers with additional data that became available after the original studies\autocite{frank05age-specific}. My analysis showed that, in each case, the groups carrying the inherited predisposing mutation had a pattern of incidence that changed relative to the control groups by reducing the estimated value of $\Gb$ by approximately one. Thus, the genetic knockouts reduced time's curvature by almost exactly the amount predicted by the reduced force of diminished simultaneity in the protective mechanisms. 

\section*{Conclusion}

A few simple invariances shape the patterns of death. That geometry does not tell us exactly how biological mechanisms influence mortality. But the geometry does set the constraints within which we must analyze the relation between pattern and process. 

I started with the temporal frame of reference, $\dT$, on which mortality has a constant rate, or velocity. That temporal frame, with unchanging rate, expresses the ticking of mortality's clock in the absence of any apparent force that would change velocity. 

Given that frame without apparent force, we can then evaluate other temporal scales in terms of the forces that must be applied to change mortality's rate relative to the force-free scale. That approach focuses attention on the forces of mortality, rather than the incidence or ``motion'' alone, because the pattern of motion is inherently confounded with the particular temporal frame of reference\autocite{lanczos86the-variational,frank15dalemberts}. 

Mortality's temporal frame leads to a natural expression of invariant death with respect to a universal Gompertzian geometry. That geometric expression separates the uniform application of force from the additional distortions of time with respect to observed pattern. 

The examples of nematodes and cancer illustrated how to parse observable deformations of mortality's clock with respect to invariant aspects of pattern and potential underlying explanations about process.

Until biologists can see the constraints of Gompertzian geometry on the curves of death as clearly as they can see the constraints of the logarithmic spiral on the growth curves of snail shells and goats' horns, we will not be able to read properly the relations between the molecular causes of failure and the observable patterns of death.

Put another way, geometry does not tell one how to build a bridge. But one would not want to build a bridge without understanding the constraints of geometry. Properly interpreting the duality of constraint and process with respect to pattern is among the most difficult and most important aspects of science.

\subsection*{Author contributions}
SAF did all the research and wrote the article.

\subsection*{Competing interests}
No competing interests were disclosed.

\subsection*{Grant information}
National Science Foundation grant DEB--1251035 supports my research.


{\small\bibliographystyle{unsrtnat}
\bibliography{mortality}}

\end{document}